\documentclass[journal=nalefd,manuscript=article]{achemso}

\usepackage[version=3]{mhchem} 

\usepackage{graphicx}
\usepackage{sidecap}
\usepackage{amsmath}
\usepackage{bm}

\author{Nicol$\grave{\rm o}$ Accanto}
\email{nicolo.accanto@icfo.es}

\author{Jana B. Nieder}

\author{Lukasz Piatkowski}

\author{Marta Castro-Lopez}

\author{Francesco Pastorelli}

\author{Daan Brinks}

\author{Niek F. van Hulst}
\affiliation{ICFO - Institut de Ciencies Fotoniques, Mediterranean Technology Park, 
08860 Castelldefels (Barcelona), Spain}

\alsoaffiliation{ICREA - Institucio Catalana de Recerca i Estudis Avancats (ICREA), 08010 Barcelona, Spain}

\title[\texttt{achemso} demonstration]
{Second harmonic nano-particles for femtosecond coherent control on the nanoscale}

\begin{document}
\date{\today}

\begin{abstract}
We provide a complete toolkit for coherent control experiments on the nano-scale. By exploiting the second harmonic emission from single (150\,nm) nonlinear nano-particles, we show that ultrafast femtosecond laser pulses can be compressed and controlled in time with unprecedented spatial accuracy. The method is tested on various nano-particles of different sizes, shapes and materials, both dielectric $\textnormal{BaTiO}_3$, $\textnormal{Fe(IO}_3\textnormal{)}_3$)  and metallic (Au) thus demonstrating its robustness and versatility.
\end{abstract}

\section{Introduction}

Investigation of the fundamental dynamic processes in individual nano-objects, such as coherent dynamics in photosynthetic complexes and femtosecond exciton dephasing in quantum dots (QDs), requires a combination of femtosecond excitation and nanometer spatial resolution. In recent years, thanks to advances in femtosecond lasers, microscopy and nano-technology, the first experimental advances in this nanometer/femtosecond domain have been realized \cite{BrinksNature2010,Aeschlimann:2011bj,Anderson:2010do,Schmidt:2012hh,Biagioni:2012hr}.

In a typical ultrafast coherent control experiment one excites a system with a defined sequence of phase locked pulses and monitors the variation of its optical response as a function of the time delay between the pulses and/or their relative phases. 
These experiments therefore rely on the ability of accurately shaping a laser pulse in the time and phase domain to match the characteristics of the particular system under study 
\cite{Warren:1993tl}. 
In the nanometer regime coherent control becomes particularly complicated for two main reasons: firstly femtosecond pulses get easily distorted while propagating through dispersive elements, for instance lenses and high numerical aperture (NA) objectives. In the time domain such phase distortions result in a stretched pulse, i.e. longer than the Fourier limit \cite{Fuchs:2005uw}. This imposes the need of a method for compensating for such distortions and re-compressing the pulse in time. Once the phase distortions are accurately corrected it is possible to shape the laser pulse introducing a known spectral phase.
Secondly, the system under study is much smaller than the wavelength of light. Practically, this means that one ideally needs to either use sub-diffraction limited (i.e. near field) field confinement at the excitation, or make sure that the temporal dynamics of the pulse are uniform over the focal volume. Often however, this is hard to achieve and the temporal pulse dynamics are inhomogeneous over the pulse focus \cite{Brinks:2011ta,McCabe:2011gx}. To either check the focus homogeneity or to deal with the inhomogenity, the ability to probe and control the femtosecond dynamics of a laser pulse in a diffraction limited focus  at sub-diffraction limited length scales is therefore necessary.
Moreover, studying systems smaller than the wavelength of light calls into question the fundamental notion of phase and to which extent phase can be defined at those length scales. Using a broadband laser source whose spectral phase can be calibrated on a nanometer spatial scale, such a phase investigation can in principle be carried out.
To get access to nanometer/femtosecond coherent control as well as fundamental phase investigation on the nano-scale therefore a method is required for compressing and controlling a laser pulse with spatial resolution beyond the diffraction limit.

The second harmonic (SH) emitted by a non-linear crystal, owing to its intrinsic coherence, is sensitive to the spectral phase of a laser pulse and therefore has been widely used for femtosecond pulse characterization, mainly in combination with Frequency resolved optical gating (FROG) \cite{Trebino:2011dya} or Spectral phase interferometry for direct electric-field reconstruction of ultrashort optical pulses (SPIDER) \cite{Iaconis:98}, and for both pulse characterization and compression with the so called Multiphoton Intrapulse Interference Phase Scan (MIIPS) technique \cite{Coello:2008vu,Xu:2006tq}.

In this work we have used MIIPS as a method for measuring and compensating for the phase distortions of a laser pulse after passing through a high NA microscope objective. In its standard implementation MIIPS makes use of a bulk SH crystal and a pulse shaper for retrieving the spectral phase of a laser pulse and correcting for the phase distortions introduced by the optics. MIIPS is an excellent tool for addressing the femtosecond part of the problem, but the spatial resolution the method can achieve is intrinsically limited when collecting the SH from a bulk crystal.
Efficient pulse compression using MIIPS has been demonstrated in the focus of a high NA objective for ultrafast microscopy applications \cite{Pestov:2009td}. However, even in the best case, the spatial resolution is limited by the size of the focused beam at the sample position, that is the diffraction limit. Inhomogeneities in the focal volume cannot be addressed by standard MIIPS as doing so would require beating the diffraction limit for light. 

To overcome this limitation and obtain a higher spatial resolution we have used single non-linear nano-particles (NPs), instead of bulk crystals, as nanometer sources of SH generation. 
Because NPs are smaller than the excitation wavelength, phase matching conditions do not apply \cite{LeXuan:2008bl}. As a consequence, efficient SH generation can be achieved from very broad wavelength ranges at the same time without the need of specifically sizing the NPs nor that of changing their orientation with respect to the incident light. This makes them more cost effective and easier to handle than their bulk crystal counterpart.

Recently the SH emission from single non-linear NPs interacting with ultrafast pulses has been detected and studied \cite{Extermann:2008bm,Li:2010hz, Wnuk:2009tx, Aulbach:2012gp, umner:2010tg,Staedler:2012bn}. Extremann \textit{et al.}  \cite{Extermann:2008bm} used single SH NPs to characterize broadband pulses with high spatial resolution using FROG; Wnuk \textit{et al.} \cite{Wnuk:2009tx} compared the SH response from NPs excited with chirped 200\,fs and compressed 13\,fs pulses; Aulbach \textit{et al.} \cite{Aulbach:2012gp} used the SH from NPs as a feedback signal for spatio-temporal focusing. 

In this paper we go further and we demonstrate full femtosecond pulse control on the nano-scale. 
First, we show full compression of a 17\,fs pulse with MIIPS in the focus of a high NA objective using the SH emitted by a single NP as small as $\sim 150$\,nm. Once the pulse has been compressed we also show deterministic control of the SH spectrum of single NPs depending on the applied spectral phase. In order to verify the robustness and reproducibility of the method we performed the same experiments on particles of different size, shape and material (both dielectric and metallic) and always found good agreement among the results. 
These findings unambiguously prove our full control of the temporal shape of a femtosecond pulse with nanometer spatial accuracy.
At the same time, they provide a complete toolkit for setting up coherent control experiments and phase investigation on the nano-scale that can be readily implemented in a variety of microscope setups.

\section{Experiment and samples}

The experimental setup that we used consists of a broadband Titanium Sapphire laser (Octavius 85 M, Menlo Systems) which was tuned to a central wavelength of $\sim 800$\,nm with a bandwidth of $\sim 90$\,nm calculated from the points where the intensity of the spectrum dropped to the 30\% of its maximum value. The repetition rate of the laser is 85\,MHz and the average power at the sample position was varied from 1 to 6\,mW.
The laser was sent through a pulse-shaper based on a Liquid-Crystal (LC) Spatial Light Modulator (SLM) integrated in a 4f-configuration, into an inverted confocal microscope. The laser beam was focused to a diffraction limited spot with a 1.3 NA objective (Fluar, Zeiss) onto a thin ($150\mu$m) microscope cover slip containing the sample, namely the SH NPs. A nanometer precision piezoelectric scanner (MadCityLabs) allowed to scan the sample area over several microns. The SH from the NPs was collected in reflection through the same 1.3 NA objective and sent either to an Avalanche photo-diode (APD, Perkin-Elmer) for imaging purposes or to a spectrograph (Shamrock SR-303i) and detected with an electron-multiplying charge-coupled device (emCCD) camera (Andor, iXon) for spectral measurements. 

\begin{figure}[!h]
\centering
\includegraphics*[width=0.8\columnwidth]{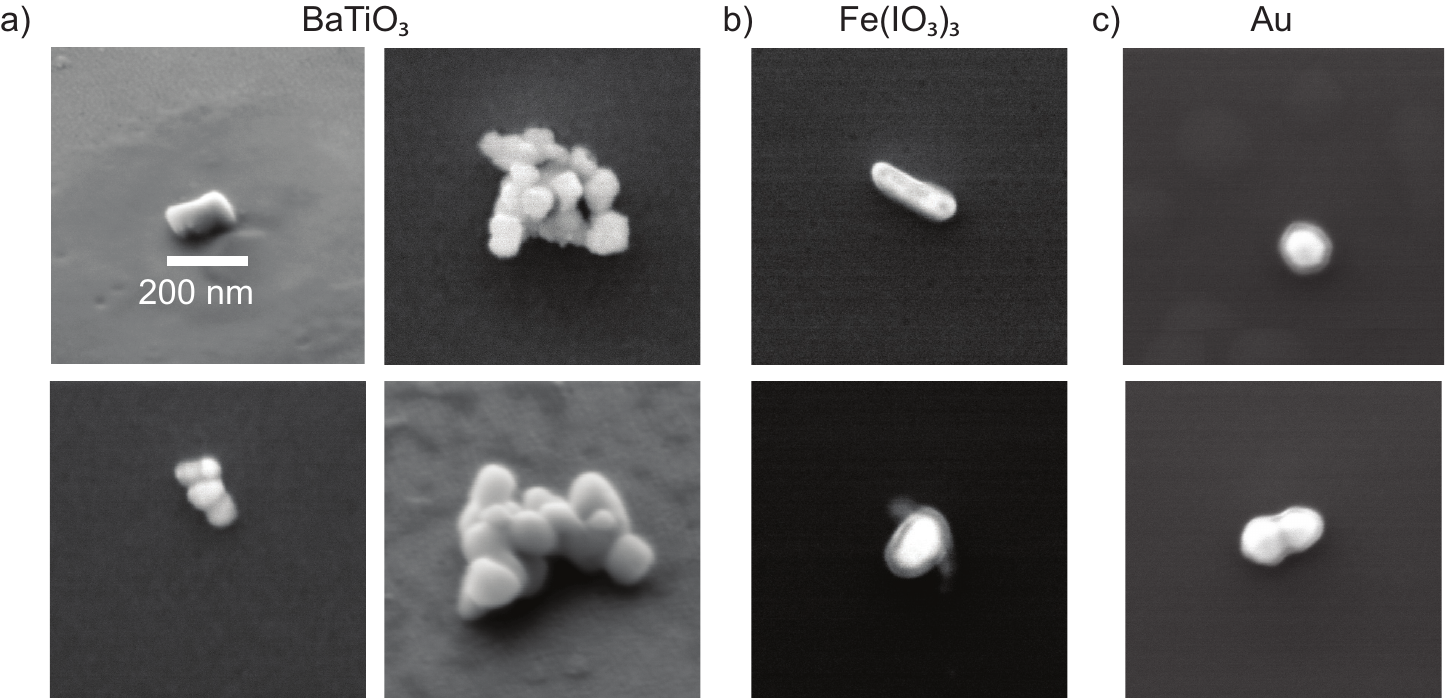}
\caption{SEM images of typical NPs we investigated. a) $\textnormal{BaTiO}_3$ NPs, b) $\textnormal{Fe(IO}_3\textnormal{)}_3$ NPs, c) Au NPs. The scale is the same in all images} 
\label{FigSEM}
\end{figure}

We studied three different types of SH NPs: barium titanate ($\textnormal{BaTiO}_3$), iron iodate ($\textnormal{Fe(IO}_3\textnormal{)}_3$) and gold (Au) nano-particles, as illustrated by the SEM images in Fig. \ref{FigSEM}. 
$\textnormal{BaTiO}_3$ (Sigma Aldrich) and $\textnormal{Fe(IO}_3\textnormal{)}_3$ (Plasmachem) were dispersed in ethanol, sonicated and finally deposited onto a glass microscope cover slip and let dry. 
$\textnormal{Fe(IO}_3\textnormal{)}_3$ NPs have mean dimensions of $\sim 150$\,nm and were homogeneously distributed on the coverslip, while in the case of $\textnormal{BaTiO}_3$ an inhomogeneous distribution of single NPs and small aggregates with sizes ranging from 100 to few hundreds nanometer was found. This makes the  $\textnormal{BaTiO}_3$ sample very suitable for size- and shape-dependent experiments.

Spherical Au NPs of 100\,nm diameter, suspended in water solution and citrate-coated, were purchased from BB International (UK) and prepared by interchanging the coating from citrate to BSPP following the method described in Ref. \cite{Bidault:2012uj}(50mL of citrate-coated particles are left to incubate overnight with 10mg BSPP). The particles were then centrifuged to remove the supernatant before being rinsed and finally resuspended in a 0.5mM BSPP solution. The final Au NP concentration is of the order of $0.02\mu \textnormal{M}$. The Au dimer shown in the bottom right image of Fig.  \ref{FigSEM} is formed by aggregation without any linker and, therefore, the two Au NPs are in direct contact.

The experiment was performed as follows: first, optical images of the NPs were acquired scanning the sample with the piezoelectric scanner and collecting the SH signal with the APD. Next, while continuously exciting one NP and recording its SH spectrum, we performed MIIPS in order to measure the spectral phase and compress the initially distorted pulse down to its Fourier limit.

During a MIIPS iteration a set of reference phase functions is applied on the SLM and the full SH spectrum is detected, which allows to retrieve the initial spectral phase. In a second step the negative of the retrieved phase (which is called a compensation mask) is applied on the SLM in order to correct for the distortions and a new MIIPS iteration is run which measures the new spectral phase and creates a new compensation mask. We repeat this process until a compensation mask that yields a constant spectral phase over the full laser spectrum (i.e. is producing a Fourier limited pulse) is found. For details on how MIIPS works see Ref.  \cite{Coello:2008vu} and references therein.  

For all the reported simulated SH spectra in the paper the software $femto\textnormal{Pulse Master}^{\textnormal{TM}}$ V1.1 from BioPhotonic Solutions Inc. was used.

Scanning electron microscopy (SEM) was used to characterize the size and shape of the NPs studied in the optical measurements. 

\section{Results and discussion}

\begin{figure}[!h]
\centering
\includegraphics*[width=\columnwidth]{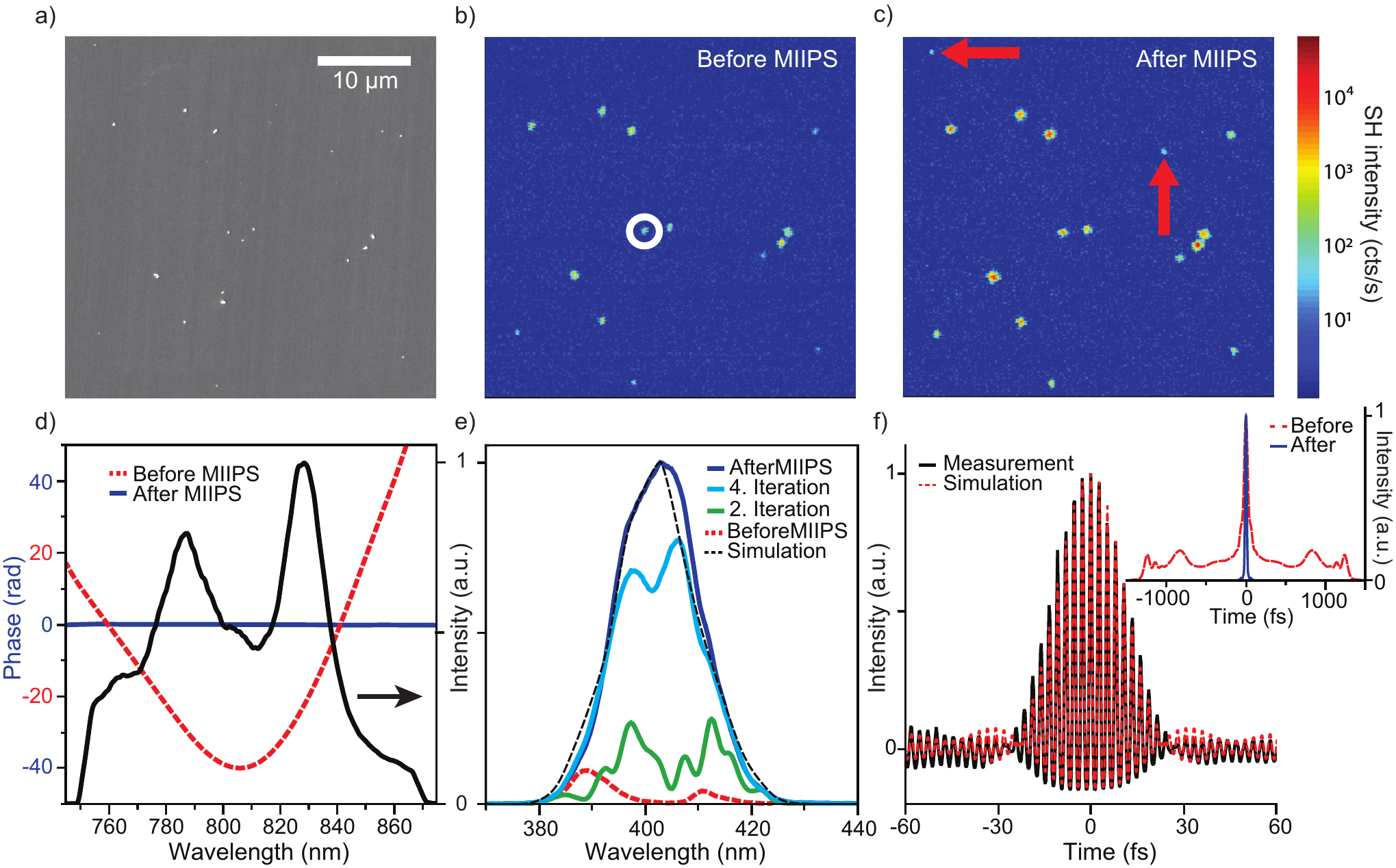}
\caption{Femtosecond pulse compression on a single 180\,nm  $\textnormal{BaTiO}_3$ NP. \protect \\
a) SEM image of the $\textnormal{BaTiO}_3$ sample area investigated. b) SH image of the sample area before pulse compression. The circled particle was chosen for pulse compression. c) SH image of the sample area after pulse compression measured with identical acquisition settings. Note the increase of the total SH collected from the NPs (same colorbar applies for both SH images). Red arrows point at small NPs (125\, nm mean dimension) which are clearly resolved only after pulse compression. d) Laser spectrum used in the experiment together with the spectral phase measured before and after MIIPS. Note that MIIPS was successfully compensating for $\sim 40$\,rad distortions. e) SH spectra of the selected NP as measured for different compensation masks during MIIPS optimization. Note that the SH spectrum after MIIPS optimization is agreeing very well with the simulated one. f) Simulated (red dashed line) and measured interferometric autocorrelation (black line) of the compressed 17\, fs pulse, as explained in the text. Inset: calculated time profile of the uncompressed (dashed red line) and compressed pulse (solid blue line).} 
\label{FigMIIPS}
\end{figure}

\begin{figure}[!h]
\centering
\includegraphics*[width=\columnwidth]{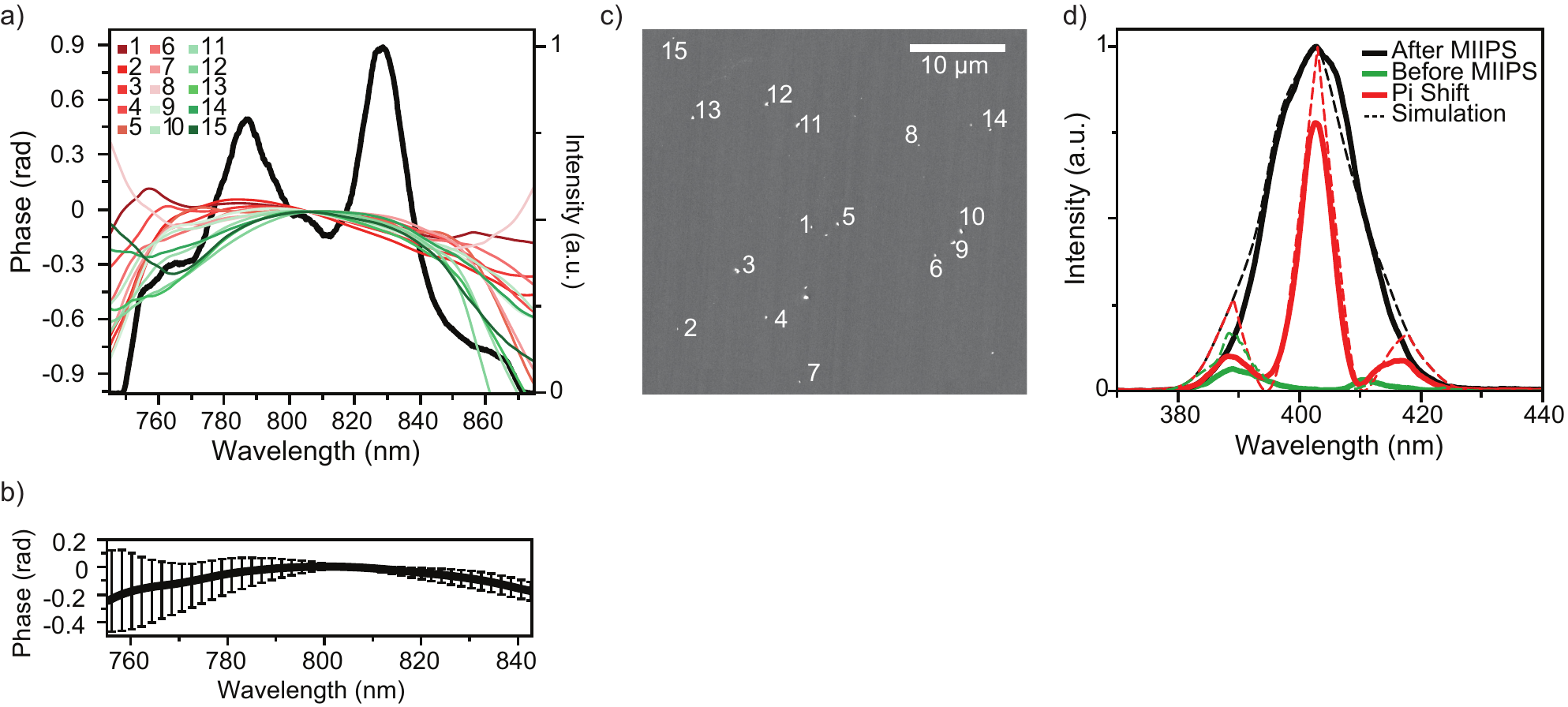}
\caption{a) Laser spectrum and spectral phase measured on different $\textnormal{BaTiO}_3$ NPs. b) Mean spectral phase over the 15 measurements. The error bars are given by the standard deviation on the 15 measurements. Note that the total variation of the spectral phase is lower than 0.3\,rad and the standard deviation is about 0.15\,rad. c) SEM image of the sample showing the 15 $\textnormal{BaTiO}_3$ NPs we measured on. d) SH control on particle 8. SH spectra measured before, after compression and for a step function of amplitude $\pi$ applied on the SLM are compared with the calculated spectra.} 
\label{FigPhaseStab}
\end{figure}

\begin{figure}[!h]
\centering
\includegraphics*[width=\columnwidth]{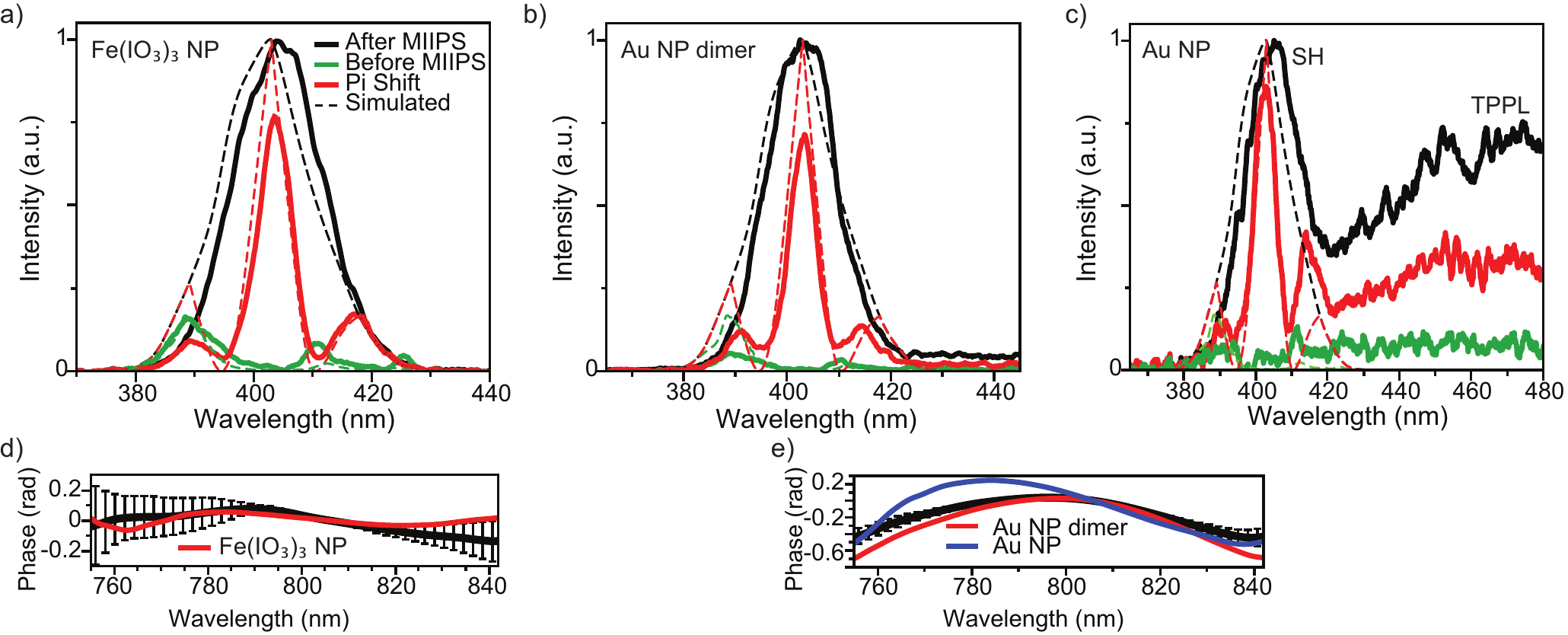}
\caption{Top: SH control on a $\textnormal{Fe(IO}_3\textnormal{)}_3$ NP a), on a Au dimer b) and on a single Au NP c). For each NP the SH spectra were collected for compressed (black), uncompressed pulses (green) and a step function of amplitude $\pi$ (red) applied to the SLM, and are shown together with the corresponding simulated SH spectra (dashed lines). For the single Au NP the wavelength range was increased up to 480 nm to include the TPPL emission.  
Bottom: spectral phase measured for one d) $\textnormal{Fe(IO}_3\textnormal{)}_3$ NP (red line), e) Au dimer (red line) and Au monomer (blue line) when applying a compensation mask obtained on a $\textnormal{BaTiO}_3$ NP. The error bars are the standard deviation obtained from repeated measurements on $\textnormal{BaTiO}_3$ NPS.}
\label{SHControl}
\end{figure}

We first studied the SH response of $\textnormal{BaTiO}_3$ NPs which we take to be our reference sample (Fig. \ref{FigMIIPS} and \ref{FigPhaseStab}). We then tested our method on $\textnormal{Fe(IO}_3\textnormal{)}_3$ and Au NPs to verify its robustness and versatility (Fig. \ref{SHControl}). 

In Fig. \ref{FigMIIPS} the results of the pulse characterization and compression using the SH signal from a single $\textnormal{BaTiO}_3$ NP are reported. The upper panel shows the SEM image (\ref{FigMIIPS}a) and optical images of the same sample area, as measured before (\ref{FigMIIPS}b) and after (\ref{FigMIIPS}c) pulse compression under otherwise identical experimental conditions. SH emission from single NPs is clearly resolved in the optical images. More than one order of magnitude increase in the total SH signal from NPs is observed upon excitation with compressed pulses. The NP highlighted with a circle in Fig. \ref{FigMIIPS}, which with SEM was measured to have a mean dimension of 180\,nm, was selected for performing pulse compression. 

In Fig. \ref{FigMIIPS}d the fundamental laser spectrum used in the experiments is plotted together with the spectral phase that MIIPS retieved before and after pulse compression.
The initial spectral phase was measured to span about 40\,rad across the entire laser spectrum, reflecting the amount of phase distortion introduced by the optics we used in the experiment. After few MIIPS iterations we were able to obtain a flat spectral phase with less than $0.3$\,rad variation (see Fig. \ref{FigMIIPS}d and \ref{FigPhaseStab}b), corresponding to nearly Fourier limited pulses. 

Fig. \ref{FigMIIPS}e shows SH spectra measured on the selected NP after successive MIIPS iterations (i.e. with successive compensation masks applied on the SLM). Each MIIPS iteration is associated to an overall SH intensity increase.The initial low intensity SH spectrum (red dashed line) is a direct consequence of the high phase distortions initially present on the laser pulse and can intuitively be understood as follows: high phase distortions correspond to a stretched pulse in the time domain where different frequency components of the laser spectrum arrive at the sample at different times. Since sum frequency generation of different components cannot occur if they are not reaching the sample at the same time, the contribution of this non-linear effect to the SH signal is lower for a stretched pulse. As MIIPS proceeds and the pulse is re-compressed in time the SH intensity increases and the spectrum approaches the simulated one corresponding to Fourier limited pulses (black dashed line). The total SH intensity collected from the NP was found to increase by a factor of 15 between the uncompressed and the Fourier limited pulse (blue line in the plot). This increase is of fundamental importance for detecting even smaller particles, as indicated in Fig. \ref{FigMIIPS}c where the two NPs indicated with red arrows, that are barely visible using uncompressed pulses, become visible upon excitation with a Fourier limited pulse. As the SEM images in the right panel of Fig. \ref{FigSEM}a show, these particles have mean dimensions of 125\,nm. 

This is to our knowledge the first demonstration of femtosecond pulse compression using the SH produced by a NP of a size smaller than the diffraction limited focal spot. We stress here that the amount of SH signal we could collect from a single $\textnormal{BaTiO}_3$ NP in our experiment is enough to acquire a SH spectrum in less than 1\,sec which allows MIIPS to proceed fast. Typically less than 10 MIIPS iterations are needed to compensate for the phase distortions, when using 32 points (i.e. we acquire 32 different SH spectra) for the first 4 iterations and 128 for the rest. Full pulse compression therefore takes less then 15 minutes.

Finally to further verify the Fourier limited character of the compressed pulse, we acquired interferometric autocorrelation of the laser pulse using the SH emitted by the same NP. Such an interferometric autocorrelation trace is measured by scanning the relative time delay of two identical replicas of the original pulse generated by the pulse shaper, as described in Ref. \cite{Pestov:2009td}. The graph in Fig. \ref{FigMIIPS}f shows the integrated SH intensity as a function of the time delay between the pulse replicas. The interferometric autocorrelation function has a  FWHM of $23.6$\,fs which, for a gaussian pulse, corresponds to a pulse FWHM of $16.7$\,fs. For visualizing the dramatic effect of the compression on the time profile of the laser pulse we also plotted in the inset of the graph the calculated intensity autocorrelation (see Ref. \cite{Pestov:2009td}) for the pulse before and after compression. For the compressed pulse a single narrow peak of $23.6$\,fs FWHM is observed, while in the case of the uncompressed pulse one sees a much broader central peak (56\,fs FWHM) superimposed on a very broad pedestal containing the 85\% of the total integrated intensity and extending up to $\pm 1$\,ps from the central peak.

In order to test the reproducibility of the method and the independence on the specific particle, we measured the spectral phase on several NPs differing in size, shape and even material.  
First, we studied the SH emitted by the 15 different $\textnormal{BaTiO}_3$ NPs indicated in the SEM image of Fig. \ref{FigPhaseStab}. These NPs provide a large distribution of sizes, varying from 130 to 500\,nm mean dimension, and shapes (see representative examples in Fig. \ref{FigSEM}a.)

In Fig. \ref{FigPhaseStab}a the spectral phases measured with MIIPS on the 15 different $\textnormal{BaTiO}_3$ NPs when applying the best compensation mask previously obtained (namely the one corresponding to Fourier limited pulses) using NP no.1 as SH source are reported. A very good agreement among all the measured spectral phases was found. If one considers only the part of the fundamental spectrum where the intensity is above the 30\% of the maximum, and therefore where the measurement is the most reliable, then the mean phase varies less than 0.3\,rad and the standard deviation is below 0.15\,rad (Fig. \ref{FigPhaseStab}b), thus confirming the good level of compensation we are able to obtain and the independence of our method on the size and shape of the NP. 

In a typical pulse shaping experiment \cite{BrinksNature2010}, once a pulse has been compressed, the SLM is used to control the pulses via application of specific spectral phases, e.g. to obtain chirped pulses or a sequence of pulses with defined delays and phase relations. To demonstrate full pulse control on the nanoscale level we applied different phase masks on the SLM, acquired corresponding SH spectra and compared them with simulations. The test phases we used are the flat phase (corresponding to Fourier limited pulses), the phase before pulse compression (corresponding to distorted pulses) and a mask with a $\pi$ phase difference between the short and long wavelength parts of the laser spectrum.
Fig. \ref{FigPhaseStab}d shows the SH spectra collected on particle number 8 in the SEM images (160\,nm mean size) corresponding to such applied phases together with calculated spectra. Very good agreement is found which indicates that our laser pulse can efficiently be controlled with $\sim 100$\,nm precision. On the other hand these experiments show that the SH emission from a single NP can be deterministically manipulated in its frequency domain. In practice, once our compression method is applied, these NPs can be considered as nanometer sources of easily tunable, coherent blue light, with pulse lengths and wavelengths all controlled by application of spectral phase.

As a final test of reproducibility and versatility of our method, we performed similar measurements on particles made of $\textnormal{Fe(IO}_3\textnormal{)}_3$ and Au. The results for these NPs are reported in Fig. \ref{SHControl}.
The main objective of these experiments is to verify that the same spectral phase is measured regardless the material the SH NPs are made of. 

For this we proceeded as follows: first we compressed the laser pulse inside the focus of the microscope objective using $\textnormal{BaTiO}_3$ NPs, retrieving a phase varying less than 0.3\,rad over the laser spectrum, as described above. Then, a few phase measurements were performed on different $\textnormal{BaTiO}_3$ NPs in order to get the mean phase and standard deviation. At this point we changed either to the $\textnormal{Fe(IO}_3\textnormal{)}_3$ or the Au sample and ran a phase measurement on one NP starting from the compensation mask obtained from MIIPS on $\textnormal{BaTiO}_3$ NPs.
In Fig. \ref{SHControl} such phase measurements together with SH spectra for different applied phases are shown for one $\textnormal{Fe(IO}_3\textnormal{)}_3$ NP of 110 by 160\,nm in size (see Fig. \ref{FigSEM}b bottom), and two different Au NPs: a dimer constituted of two attached 100\,nm Au spheres and a single 100\,nm Au sphere (see Fig. \ref{FigSEM}c). For these particles the laser power was reduced from 6\,mW to about 1\,mW to avoid photo damage. 
Even applying this lower power to $\textnormal{Fe(IO}_3\textnormal{)}_3$ we were able to collect SH spectra with lower than 1\,sec integration time for 150\,nm particles, confirming the good SH efficiency of this material \cite{Extermann:2008bm}. For Au NPs shorter than 1\,sec integration time was still possible in the case of the Au dimer, while for the single 100\,nm Au NPs we needed to increase the integration time to 2\,sec or longer to have an adequate signal to noise ratio in the SH spectra.

As illustrated Fig. \ref{SHControl}d, for $\textnormal{Fe(IO}_3\textnormal{)}_3$ NPs the measured phase is well within the standard deviation obtained previously on $\textnormal{BaTiO}_3$ NPs. Moreover the SH spectra acquired before compression, after compression and with a $\pi$ amplitude step function are again largely matching the simulations Fig. \ref{SHControl}a. 

Next we turn to the study of Au NPs.
Au NPs are known to undergo both SH generation \cite{Butet:2010es, Slablab:2011tl} and two-photon absorption giving rise to two photon photoluminescence (TPPL) \cite{CastroLopez:2011koa}. The onset of the TPPL emission spectrally overlaps with the long wavelength part of the SH spectrum. TPPL from Au has been successfully used for mapping the plasmonic resonances of Au NPs \cite{Ghenuche:2008ja}. However in the present case high TPPL is rather a source of noise which might disturb the precise measurement of the SH spectrum and therefore of the spectral phase. 
In qualitative agreement with a recent work by Deng at al. \cite{Deng:2013cu} we found that the ratio between SH and TPPL is larger for Au dimers than for single 100\,nm Au NPs. Moreover the TPPL emission, as expected for a non-linear process, changes as a function of the spectral phase, being maximal for fully compressed laser pulses.
As shown in Fig. \ref{SHControl}c for single Au NP no clear SH could be measured with the typical acquisition settings without compressing the laser pulse. In order to have a good signal to noise ratio for the SH signal at least 5\,sec integration time was needed. 
Nevertheless even for this particle, regardless the long integration time and the very high TPPL contribution partly overlapping with the SH signal, remarkably the spectral phase MIIPS measured was no more than 0.2\,rad different from that obtained on $\textnormal{BaTiO}_3$ NPs and on the Au dimer (see Fig. \ref{SHControl}e). Moreover the SH spectrum could still be accurately controlled in the case of the Au dimer (Fig. \ref{SHControl}b) and to a good extent on the single NP as well (Fig. \ref{SHControl}c). 

These observations lead to two important conclusions: on one hand they poof the independence of the phase measurement and SH control on the material the NPs are made of, which further supports the robustness of our method; on the other hand they state that the phase response of all the NPs we investigated is the same, i.e. there is no additional phase information associated to the NPs.
A priori this is to be expected for dielectric materials such as $\textnormal{BaTiO}_3$ and $\textnormal{Fe(IO}_3\textnormal{)}_3$, that are transparent in our fundamental laser spectrum, while  it is not necessarily the case for Au NPs.
Indeed, it is well known that metallic NPs can show plasmonic resonances in the visible and near infrared spectrum \cite{Novotny:2011ko}. Such resonances carry a phase information which in our experiment might show up as additional features in the phase measurement and therefore might affect the measurement of the laser spectral phase. 
Since our objective here was the precise measurement of the laser spectral phase, we chose the Au NPs to be off-resonant with the excitation, as confirmed from extinction measurements (not shown in the paper) which show the plasmonic resonance centered at $\sim 570$\,nm for the investigated NPs.

\section{Conclusions and Outlook}

In this paper we have shown that the second harmonic from individual nano-particles allows to obtain full spatial and temporal control of a laser pulse with nanometer and femtosecond accuracy.
We have demonstrated this ability in two steps: first, using the SH from a single 150\,nm NP we were able to fully compress a femtosecond laser pulse in the time domain down to its Fourier limit using MIIPS; second, we have shown that it is possible to deterministically control the SH emission from a single NP when applying specific spectral phases to the laser pulse. We finally tested the robustness of our method by performing the same experiments on particles of different sizes, shapes and materials. 

These results together demonstrate full femtosecond pulse control on a nanometer spatial scale and pave the way towards coherent control experiments on the nanoscale. 
In addition, the accuracy of phase control at the nanoscale that we achieved can enable the study of the phase response in resonant plasmonic NPs.
Finally, our SH control allows SH NPs to be used as nanometer sources of tunable  coherent blue light. 

\acknowledgement

This research was funded by the MICINN (programs Consolider Ingenio-2010: CSD2007-046 - NanoLight.es, Plan Nacional FIS2009-0123: Optical NanoAntennas) and the European Union (ERC Advanced Grant 247330- NanoAntennas). L.P. acknowledges financial support from Marie-Curie International Fellowship COFUND and ICFOnest program, F.P. received support from the European Commission through  the Erasmus Mundus Joint Doctorate Programme Europhotonics (Grant No. 159224-1-2009-1-FR-ERA MUNDUS-EMJD). D.B. acknowledges support by a Rubicon Grant of the Netherlands Organization for Scientific Research (NWO)

\bibliography{Nanocrystals_biblio}

\end{document}